\documentclass[aps,prd,preprint,superscriptaddress,showpacs,nofootinbib]{revtex4-1}
\usepackage{float}
\usepackage{graphicx}
\usepackage{graphics}
\usepackage{epsfig}
\usepackage{latexsym}
\usepackage{colordvi}
\usepackage{amsmath}
\usepackage{amssymb}
\usepackage{slashed}
\usepackage{xcolor}

\usepackage[title]{appendix}

\newcommand{\dis}[1]{\begin{equation}\begin{split}#1\end{split}\end{equation}}

\begin{document}

\preprint{\parbox[b]{1in}{ \hbox{\tt PNUTP-22/A04}  }}
\preprint{\parbox[b]{1in}{ \hbox{\tt CTPU-PTC-22-16}  }}

\title{Detecting axion dark matter with chiral magnetic effects}
\author{Deog Ki Hong}
\email[E-mail: ]{dkhong@pusan.ac.kr}
\affiliation{Department of Physics, Pusan National University,
             Busan 46241, Korea}
\author{Sang Hui Im}
\email{imsanghui@ibs.re.kr}            
\affiliation{Center for Theoretical Physics of the Universe,
Institute for Basic Science, Daejeon 34126, Korea}
\author{Kwang Sik Jeong}
\email[E-mail: ]{ksjeong@pusan.ac.kr}
\affiliation{Department of Physics, Pusan National University,
             Busan 46241, Korea}
             

\author{Dong-han Yeom}
\email{innocent.yeom@gmail.com}
\affiliation{Department of Physics Education, Pusan National University, Busan 46241, Korea}             

\vspace{0.1in}

\date{\today}

\begin{abstract}
We show that dark matter axions or axion-like particles (ALP) induce spontaneously alternating electric currents in conductors along the external magnetic fields due to the (medium) axial anomaly, realizing the chiral magnetic effects (CME). We propose a new experiment to measure this current to detect the dark matter axions or ALP. These induced currents are the electron medium effects, directly proportional to the axion or ALP coupling to electrons, which depends on their microscopic physics. In the experimental setup one measures the sum of the electric current due to CME and the vacuum current due to the anomalous axion-photon coupling. The CME current is in general subdominant by a factor of the Fermi velocity of electrons, compared to latter, unless the axion or ALP coupling to electrons is much bigger than its coupling to photons to compensate the Fermi velocity suppression. However, we find that repurposing the currently operating and planned axion haloscopes may have good sensitivity to probe the CME current.

\end{abstract}


\maketitle

\newpage

\section{Introduction}
Though none of new particles beyond the standard model (BSM) have been found yet despite of tremendous experimental endeavors  for decades, 
the axion is still one of the well-motivated new particles, currently being 
searched actively in the laboratory and also in the sky~\cite{Choi:2020rgn}. 
It could explain why quantum chromodynamics (QCD) preserves the time reversal symmetry or CP and also the parity P, known as the strong CP problem. 
Furthermore it is an excellent candidate for dark matter that constitutes roughly a quarter of our universe~\cite{Chadha-Day:2021szb}. 

QCD allows in the Lagrangian a marginal theta term that breaks CP and P in general
\begin{equation}
{\cal L}_{\rm QCD}\supset \frac{\theta}{16\pi^2}\epsilon^{\mu\nu\alpha\beta}\,{\rm Tr}\,F_{\mu\nu}F_{\alpha\beta}\,,
\end{equation}
where  $F=dA+A^2$ is the field strength two form of the gluon fields $A$. 
The $\theta$ parameter, being the coefficient of the Pontryagin index,  has to be an angle variable with the $2\pi$ periodicity for the gauge theory to be consistent quantum mechanically. One of the viable solutions for the strong CP problem is to promote the $\theta$ angle to a dynamical field, namely the axion, $\theta=a(x)/f$, with the axion decay constant $f$, that relaxes to the CP-symmetric value.

Since the parameter $\theta$ shifts under the anomalous ${\rm U}(1)_A$ axial transformation of colored fermions that changes the phase of fermion mass matrix, $M$,
\begin{equation}
\theta\mapsto\bar\theta=\theta+{\rm arg}\det M\,,
\end{equation}
the axion can be realized as a pseudo Nambu-Goldstone boson of the Peccei-Quinn mechanism for the strong CP problem~\cite{Peccei:1977hh}. When QCD confines at a scale $\Lambda_{\rm QCD}\sim 250~{\rm MeV}$, the axion gets mass if the fermions are not massless~\cite{Weinberg:1977ma,Wilczek:1977pj}. The axion mass is then approximately given by  
\begin{equation}
m_a\sim \frac{\sqrt{m\Lambda_{\rm QCD}^3}}{f}\,,
\end{equation}
where $m$ is the light quark mass, $m\ll\Lambda_{\rm QCD}$. 

The experimental search in the laboratory and also the astrophysical observation strongly constrain the axion mass or its decay constant. 
For a large decay constant, axions couple sufficiently weakly to the standard model (SM) particles to constitute the main component of the energy density of our universe by the so-called misalignment mechanism~\cite{Preskill:1982cy,Abbott:1982af,Dine:1982ah,Turner:1985si,Bae:2008ue}:
\begin{equation}
\Omega_ah^2\approx 0.23\times 10^{\pm0.6}\left(\frac{f}{10^{12}~{\rm GeV}}\right)^{1.175}\theta_i^2F(\theta_i)\,,
\end{equation}
where $h$ is the Hubble expansion parameter in units of $100\,{\rm km\,s^{-1}Mpc^{-1}}$ and $\theta_i$ the initial misalignment angle with $F(\theta_i)$ being the correction due to the anharmonic effects. 

The current laboratory searches for the axion or axion-like dark matter (ADM) rely on its couplings to the SM particles
such as its anomalous coupling to photons~\cite{Sikivie:1983ip,Sikivie:2013laa,ADMX:2020ote,Semertzidis:2019gkj,Beurthey:2020yuq, HAYSTAC:2023cam, Grenet:2021vbb, McAllister:2017lkb, QUAX:2023gop, Kahn:2016aff,Obata:2018vvr,Marsh:2018dlj,Berlin:2020vrk,Schutte-Engel:2021bqm, Gramolin:2020ict, Salemi:2021gck, DMRadio:2022jfv} or its spin coupling to SM fermions~\cite{Krauss:1985ub,Stadnik:2013raa,Abel:2017rtm,JacksonKimball:2017elr,Graham:2017ivz}. In this work we propose a new experiment, ``Low temperature Axion Chiral Magnetic Effect (LACME)," to detect ADM, utilizing its coupling to the axial density of electrons in medium like a metal or an electric conductor, where the time derivative of axion field, $\dot a(t)$, plays a role of the axial chemical potential for electrons, $\mu_5$. By the chiral magnetic effect (CME)~\cite{Fukushima:2008xe} in the medium of gapless electrons spontaneous electric currents will be then generated along the external magnetic field in the presence of ADM. The CME is nothing but the helicity imbalance of fermions in gapless medium  by the axial chemical potential under the external magnetic fields~\cite{Hong:2010hi}.  If this induced (anomalous) current is measured, we will be able to determine the coupling strength of axions to electrons, which might uncover the microscopic origin of axions \cite{Choi:2021kuy}. 

\section{Chiral magnetic effect and axions}
In the chirally imbalanced medium that exhibits the gapless excitations of charged fermions, electric currents generate spontaneously along the external magnetic fields when the axial currents are anomalous, known as the chiral magnetic effect~\cite{Fukushima:2008xe}. The CME is intensively investigated in the heavy ion collisions~\cite{ALICE:2012nhw}, where the chiral imbalance is induced by the topological fluctuations of color fields, or in the condensed chiral matter~\cite{Li:2014bha}. We now show that the coherent axion or axion-like dark matter naturally induces the CME in the Fermi liquid, if coupled to ADM, providing a novel way to probe the properties of dark matter (DM) axions or ALP such as their couplings to electrons~\footnote{The time-dependent axion background will induce an effective current, that sources $\nabla\times \vec B$, under an external magnetic field even in the absence of matter, if the axion couples to photons~\cite{Ouellet:2018nfr}. We call this the vacuum effect to distinguish our medium effect from it. The vacuum effect has been studied in the ABRACADABRA cavity experiment~\cite{Kahn:2016aff}, which was merged into the DMRadio experiment~\cite{DMRadio:2022jfv}. The effect we propose in this work is the electron medium effect that exists on top of the vacuum effect.}.

Axions couple to electrons with strength $C_e/f$ directly at the tree level (DFSZ model~\cite{Zhitnitsky:1980tq,Dine:1981rt}) or induced by the heavy exotic quarks in loops (KSVZ model~\cite{Kim:1979if,Shifman:1979if})~\footnote{
See e.g. \cite{Srednicki:1985xd, Chang:1993gm, Bauer:2020jbp, Choi:2021kuy} for the discussion of the axion-electron coupling in the DFSZ model and the KSVZ model.},
\begin{equation}
{\cal L}_{\rm int}=C_e\frac{\partial_{\mu}a}{f}\bar\psi\gamma^{\mu}\gamma_5\psi\,.
\label{axion}
\end{equation}
On the other hand, if the axion is originated from a zero mode of a higher dimensional form field in string theory, $C_e$ is typically of one-loop order~\cite{Choi:2021kuy}. Depending on models, therefore, the strength of the axion-electron coupling varies as
\begin{equation}
C_e \simeq 
\begin{cases}
{\cal O}(1) & \textrm{DFSZ-like models} \\
{\cal O}(10^{-4} \sim 10^{-3}) & \textrm{KSVZ-like models} \\
{\cal O}(10^{-3} \sim 10^{-2}) & \textrm{string-theoretic axions} \,.
\end{cases} \label{Ce}
\end{equation}
The precise numerical value of $C_e$ depends on model parameters such as $f$ and/or the ratio of vacuum expectation values of two Higgs doublets.
Thus, in principle, a precise measurement of the axion-electron coupling such as our proposal can tell us which class of high energy physics underlies as a microscopic origin for the axion.

If the axions are the main component of dark matter, produced by the vacuum misalignment, they can be described as  a coherent classical field, given as
\begin{equation}
a(t)=\frac{\sqrt{2\rho_{\rm DM}}}{m_a}\sin\left(m_at\right)\,,
\label{dm}
\end{equation}
where $m_a$ is the axion mass and $\rho_{\rm DM}\approx 0.4~{\rm GeV}\,{\rm cm}^{-3}$ is the local dark matter energy density. 
By plugging Eq.~(\ref{dm}) into Eq.~(\ref{axion}), we find that the time-derivative of axion fields acts as an axial chemical potential.  Once the time-derivative of axion field is identified as the axial chemical potential, 
\begin{equation}
\mu_5=C_e\frac{\sqrt{2\rho_{\rm DM}}}{f}\,\cos\left(m_at\right)\sim0.25\times10^{-32}~{\rm GeV}\cdot\left(\frac{\rho_{\rm DM}}{0.4~{\rm GeV}\,{\rm cm}^{-3}}\right)^{1/2}\cdot\left(\frac{10^{12}~{\rm GeV}}{f/C_e}\right)\,,
\label{axion_chemical}
\end{equation}
the CME under the axion background follows.
%
%

To create the chiral magnetic effects in medium with non-vanishing axial chemical potential, we apply a magnetic field. 
The energy level of electrons under a constant magnetic field, $\vec B_{\rm ext}=B{\hat k}$, is quantized by the Landau level:
\begin{equation}
E_n(p_z)=\pm\sqrt{p_z^2+m^2+2\left|eB\right|n},
\end{equation}
where $2n=2n_r+1+\left|m_L\right|-{\rm sign}(eB)\left(m_L+2s_z\right)$ with $n_r$ being the number of radial nodes, $m_L$ and $s_z$ being the orbital angular momentum and spin along the direction of magnetic field, respectively. If we turn on the electron chemical potential, 
 electrons carrying momentum will populate until their energy reaches the chemical potential,  $E_n=\mu$, at each Landau level as long as $\mu>\sqrt{m^2+2\left|eB\right|n}$. 
 %
While the vector chemical potential shifts the ground state energy to populate the electrons up to the Fermi momentum $p_F$, the axial chemical potential shifts the momentum in the direction of spin to populate more the positive helicity states ($h=1/2$),  up to $p_F^+=p_F+\mu_5$, than the negative ones ($h=-1/2$), populated up to $p_F^-=p_F-\mu_5$, assuming $p_F>\mu_5$, since $\gamma^0\gamma_5=\Sigma\cdot\gamma$, where $2\Sigma^i=i\epsilon^{ijk}\gamma^j\gamma^k$. (See Fig.~1.) 
\begin{figure}[h]
\centering 
\includegraphics[scale=0.42]{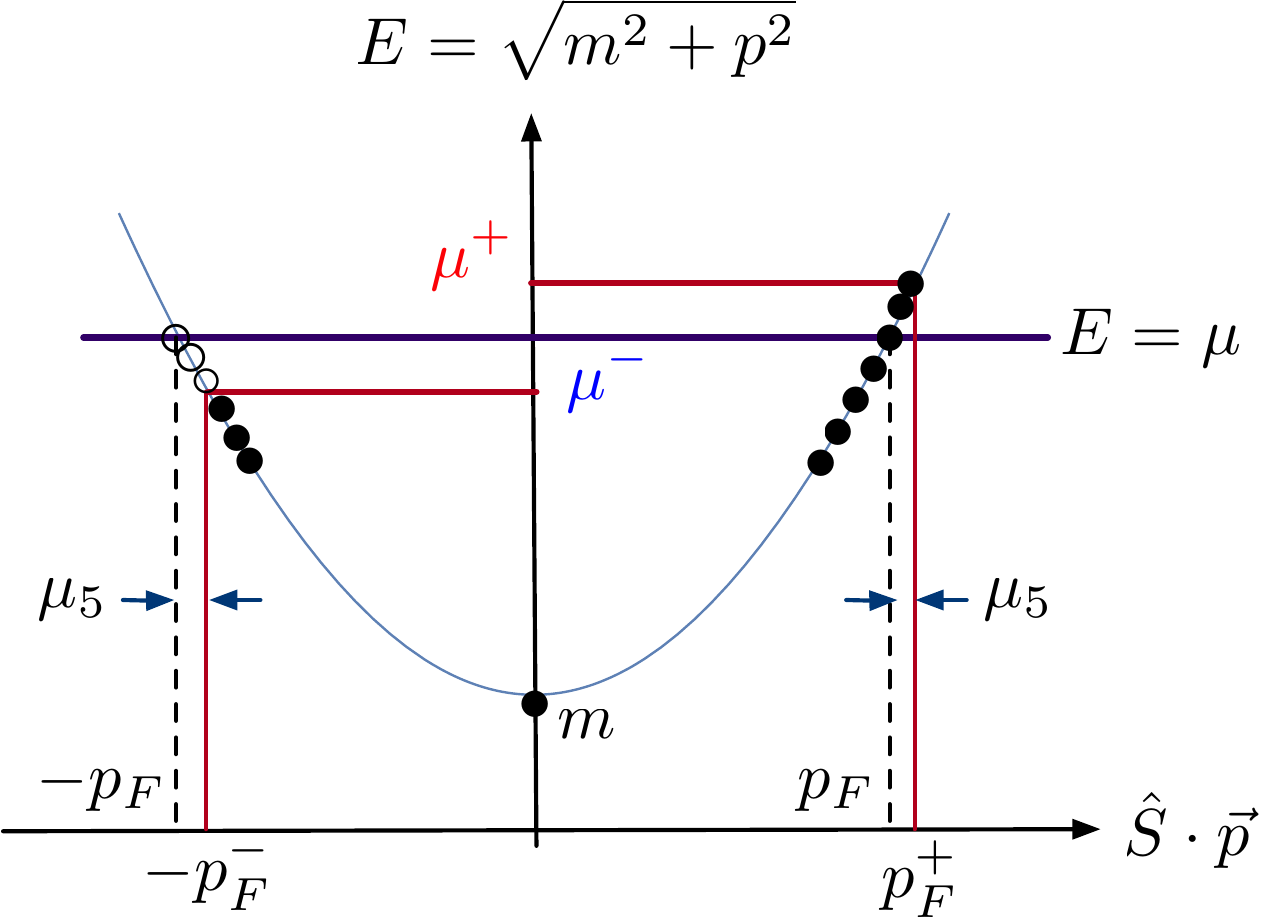}
\label{medium}
\caption{The helicity imbalance due to an axial chemical potential $\mu_5$ in dense medium. $\hat S\cdot \vec p$ denotes the momentum along the spin direction.}
\end{figure}
Namely, if we transform the electron field, $\psi\to \psi^{\prime}=e^{-i\mu_5\vec\Sigma\cdot \vec x/3}\psi$, we absorb $\mu_5$ into the momentum along the spin direction to get
\begin{equation}
{\cal L}=\bar\psi\left(i\slashed\partial -m+\mu\gamma^0+\mu_5\gamma^0\gamma_5\right)\psi=\bar\psi^{\prime}\left(i\gamma^{\prime}\cdot\partial -m+\mu\gamma^0\right)\psi^{\prime}\,,
\end{equation}
where ${\gamma^{\mu}}^{\prime}=e^{-i\mu_5\vec\Sigma\cdot \vec x/3}\gamma^{\mu}e^{i\mu_5\vec\Sigma\cdot \vec x/3}$ is a unitary transform of the $\gamma$ matrices. 
 We note, however, that the spin degeneracy is broken for the states in the lowest Landau level (LLL), having the electron spins always anti-parallel to the magnetic field~\cite{Aharonov:1978gb}. The axial chemical potential  therefore creates the helicity imbalance only for electrons in LLL to generate
 net electrons, moving antiparallel to the magnetic field or net electric current along the magnetic field, realizing the CME.

The imbalance between the helicity eigenstates, namely the difference between the number density in helicity eigenstates, which comes only from the LLL electrons\,\footnote{The spin degeneracy in the higher Landau levels, $n\ne0$, will be  slightly lifted by the Zeeman splitting due to the anomalous magnetic moment of electrons, giving rise to an additional helicity imbalance, quadratic in magnetic fields. The CME of higher Landau levels cancels out, however, because the spin up and down states contribute oppositely. Therefore, only LLL electrons contribute to the CME, unless the higher Landau levels are partially occupied. We thank Ken Van Tilburg for discussions on this.},  is given as 
\begin{equation}
	\Delta\rho=\rho^{n=0}_{h=+1/2}-\rho^{n=0}_{h=-1/2}=\frac{\left|eB\right|}{4\pi^2}\left(p_F^+-p_F^-\right)=\frac{\left|eB\right|}{2\pi^2}\mu_5\,.
\end{equation}
If we neglect the electron mass, $m=0$, all the electrons in the Fermi sea are moving with the speed of light, $c=1$, the net current due to the imbalance is then just
\begin{equation}
	\left<j^3\right>=e\,{\rm sign}(eB)\Delta\rho=\frac{e^2B}{2\pi^2}\mu_5\,.
\end{equation}
But, in the case of massive electrons, they are moving with the velocity, $p_z/E(p_z)$. The net induced current becomes therefore, summing up all the contributions from the Fermi sea,
\begin{equation}
	\!\!\!\!\left<j^3\right>=\frac{e^2B}{4\pi^2}\left[\int_0^{p_F^+}\!\!\!\!\frac{p_z{\rm d}p_z}{\sqrt{p_z^2+m^2}}
	-\int_0^{p_F^-}\!\!\!\!\frac{p_z{\rm d}p_z}{\sqrt{p_z^2+m^2}}
	\right]\!=\!\frac{e^2B}{2\pi^2} 
	\frac{2v_F\mu_5}{\sqrt{1+r^2+2v_Fr}+\!\sqrt{1+r^2-2v_Fr}}
	\,,
	\label{current}
\end{equation}
where $r=\mu_5/\mu$, while $p_F=\sqrt{\mu^2-m^2}$ and $v_F=p_F/\mu$ are the Fermi momentum and Fermi velocity of conductors when $\mu_5=0$, respectively.
We now calculate explicitly the induced (anomalous) current in helicity imbalanced medium. At one-loop the current is given by
\begin{equation}
\left<j^{\mu}\right>=e\left<\bar\Psi\gamma^{\mu}\Psi\right>=-e\int\frac{{\rm d}^4p}{(2\pi)^4}\,{\rm Tr}\left[\gamma^{\mu}S_F^{n=0}(p,\mu,\mu_5)\right]\,,
\end{equation}
where $S_F^{n=0}(p)$ is the LLL electron propagator in cold medium. Defining ${\tilde p_{\shortparallel}}=(p_0+\mu+\mu_5\gamma_5, 0,0,p_z)$ and $\vec p_{\perp}=(0,p_x,p_y,0)$, the LLL propagator in medium under an external magnetic field along $z$ direction can be written as~\cite{Gusynin:1995gt} 
\begin{equation}
	S_F^{n=0}(p,\mu,\mu_5)=\left[\frac{2i\,({\tilde {\slashed p_{\shortparallel}}}+m)P_-H_+\,e^{-p_{\perp}^2/\left|eB\right|}}{\left[\left(1\!+\!i\epsilon\right)p_0+\mu_+\right]^2-p_z^2\!-\!m^2}+\frac{2i\,({\tilde {\slashed p_{\shortparallel}}}+m)P_-H_-\,e^{-p_{\perp}^2/\left|eB\right|}}{\left[\left(1\!+\!i\epsilon\right)p_0+\mu_-\right]^2-p_z^2\!-\!m^2}\right]\,,
\end{equation}
where $\mu_{\pm}=\sqrt{{p_F^{\pm}}^2+m^2}$, the spin projection operator $P_-\!=\!(1-i\gamma^1\gamma^2\,{\rm sign}(eB))/2$ and the helicity projection operator $H_{\pm}(p)=(1\pm{\vec \Sigma}\cdot\hat p)/2$ with $\vec\Sigma=\gamma^5\gamma^0\vec\gamma$.
As the induced current vanishes in vacuum, we may write 
\begin{equation}
\left<j^{\mu}\right>=\int_0^{\mu}{\rm d}\mu^{\prime}\frac{\partial}{\partial\mu^{\prime}}	\left<j^{\mu}(\mu^{\prime})\right>\,,
\end{equation}
where $\left<j^{\mu}(\mu^{\prime})\right>$ is the induced current for the medium of chemical potential $\mu^{\prime}$. 
Following~\cite{Hong:2010hi}, we shift $p_0\to p_0^{\prime}=p_0+\mu^{\prime}$ and use $1/(x+i\epsilon)=P\frac{1}{x}-\pi i\,\delta(x)\,{\rm sign}(\epsilon)$ to get, after integrating over $\mu^{\prime}$ and taking the trace,  
\begin{eqnarray}
\left<j^{3}\right>&=&\frac{e^2B}{4\pi^2}\left[\int_0^{\mu_+}\!\!{\rm d}p_0\int_{p_z>0}\left|p_z\right|\delta\left(p_0^2-p_z^2-m^2\right)-\int_0^{\mu_-}\!\!{\rm d}p_0\int_{p_z>0}\left|p_z\right|\delta\left(p_0^2-p_z^2-m^2\right)\right]\,\nonumber\\
&=&\frac{e^2B}{4\pi^2}\,\left[\sqrt{\left(p_F+\mu_5\right)^2+m^2}-\sqrt{\left(p_F-\mu_5\right)^2+m^2}\right]=\frac{e^2B}{2\pi^2}\mu_5v_F\left[1+{\cal O}(v_F^2,r^2)\right],
\end{eqnarray}
agreeing with Eq.~(\ref{current}), where in the second line the current is doubly expanded  in powers of Fermi velocity, $v_F$, and the ratio, $r=\mu_5/\mu$. We see that CME current is generated because the axial chemical potential causes the imbalance in the helicity eigenstates, $\mu_+-\mu_-\approx 2\mu_5v_F$ at the zero temperature. This is persistent even at the finite temperature. The CME currents become at finite temperature, taking the thermal average with the Boltzmann constant $k_B=1$,  
\begin{equation}
\left<j^3\right>=\frac{e^2B}{4\pi^2}\int_0^{\infty}\frac{p_z{\rm d}p_z}{\sqrt{p_z^2+m^2}}	\left[f(p_z,\mu_+,T)-f(p_z,\mu_-,T)\right]\simeq\frac{e^2B}{2\pi^2}\mu_5v_F\left(1-e^{-(\mu-m)/T}\right),
\end{equation}
where the Fermi-Dirac distribution $f(p_z,\mu,T)=(1+\exp\{[\sqrt{p_z^2+m^2}-\mu]/T\})^{-1}$\, and we take in the last expression $v_F\mu_5\ll T\ll\mu-m$. 

We find
 that the chiral magnetic effects are realized in the electron Fermi liquid like metal. The medium dependence of CME appears only in the Fermi velocity, as the modes near the Fermi surface flip the helicity, creating a net current along the magnetic field, in the presence of the axial chemical potential. 
Interpreting as the axial chemical potential the time-derivative of the coherent axions or ALP in Eq.\,(\ref{axion_chemical}), $\mu_5=C_e \,\dot a(t)/f$, 
  the anomalous electric currents in conductors, which we call axionic CME, are given for $\vec B=B\hat k$ as
\begin{equation}
j^3=6.8\times 10^{-15}\cos\left(m_at\right)\!\cdot\left(\frac{v_F}{0.01c}\right)
\left(\frac{\rho_{\rm DM}}{0.4\,{\rm GeV}\,{\rm cm}^{-3}}\right)^{1/2}\!\!\cdot\left(\frac{10^{12}\,{\rm GeV}}{f/C_e}\right)\!\cdot\left(\frac{B}{10\,{\rm Tesla}}\right){\rm A \,m^{-2}}\,.
\end{equation}

\section{Axial anomaly and CME}
CME is the anomalous transport of electrons in the LLL due to the axial chemical potential. To see its relation with the axial anomaly, we calculate the anomalous two-point function of LLL electron currents with $\mu,\nu=0,3$,
\begin{equation}
\Gamma^{\mu\nu}(q_1)\delta^{(2)}(q_1+q_2)\equiv
\int\Pi_{i}{\rm d}^2x_ie^{iq_i\cdot x_i}
\left<0\right|{\rm T}j^{\mu}(x_1)j^{\nu}_5(x_2)\left|0\right>\,.
\end{equation}
When the medium is absent, the two-point function is given as 
\begin{equation}
	\Gamma^{\mu\nu}_{\rm vac}(q)=\frac{eB}{4\pi^2}\left(\epsilon^{\nu\alpha}q_{\alpha}q^{\mu}+\epsilon^{\mu\nu} q^2\right)H(q^2, m^2)\,,
\end{equation}
where $\epsilon^{\mu\nu}$ is the anti-symmetric tensor in (1+1) dimensions with $\epsilon^{03}=1$. The (vacuum) two-point function vanishes in the infrared, not contributing to the axial anomaly, because the loop function $H(q^2,m^2)$ does not have a pole at $q^2=0$ for massive electrons~\cite{Coleman:1982yg}. 
In medium, however, it does not vanish in the infrared because of the gapless modes at the Fermi surface. We first note that the anomalous correlators for the LLL electrons are related to the vector correlators  because of the identity ${\rm Tr}\left(\gamma^{\mu}\gamma^{\nu}\gamma_52P_-\right)={\rm sign}(eB){\rm Tr}\left(\gamma^{\mu}\gamma^{\nu}\gamma^0\gamma^3\right)$.  From the Hard Dense Loop results of the vector correlator~\cite{Manuel:1995td,Hong:1998tn} one finds the anomalous correlator to be for $q^0/\mu, \,|\vec q|/\mu\to 0$
\begin{equation}
	\Gamma^{\mu\nu}(q)=\frac{eB}{2\pi^2v_F}\left[-\eta^{\mu0}\epsilon^{\nu0}+\frac{q^0}{2}\left(\frac{V^{\mu}\epsilon^{\nu\alpha}V_{\alpha}}{V\cdot q}+\frac{{\bar V}^{\mu}\epsilon^{\nu\alpha}{\bar V}_{\alpha}}{\bar {V}\cdot q}\right)\right]\,,
	\label{ano}
\end{equation} 
where $V^{\mu}=(1,0,0,v_F)$ and ${\bar V}^{\mu}=(1,0,0,-v_F)$\,. Treating $\mu_5$ as a perturbation, we find from the anomalous two-point function, Eq.~(\ref{ano}), that the induced current under the external magnetic field becomes at the leading order in $\mu_5$
\begin{equation}
	\left<j^3\right>=-e\mu_5\underset{q_0\to0}{\rm lim}\,\underset{q_3\to0}{\rm lim}\,\Gamma^{30}(q)=\frac{e^2B}{2\pi^2}v_F\mu_5\,,
	\label{cme}
\end{equation}
which reproduces the result of CME, Eq.~(\ref{current}).\,\footnote{In a system with finite density the order of static limit and the homogeneous limit often does not commute and should be taken carefully~\cite{Feng:2020pmx}.}  We also recover the (1+1) dimensional medium axial anomaly in the background of external gauge fields $A_{\mu}$ with field strength $F_{\mu\nu}$
\begin{equation}
\left<\partial_{\nu}j^{\nu}_5\right>_A	=ie\!\int\!\frac{{\rm d}^2q}{4\pi^2}\,\underset{q_0\to0}{\rm lim}\,\underset{q_3\to0}{\rm lim}\,e^{iq\cdot x}q_{\nu}A_{\mu}(q)\Gamma^{\mu\nu}(q)= \frac{e^2B}{4\pi^2}v_F\epsilon^{\mu\nu}F_{\mu\nu}\,,
\label{anomaly}
\end{equation}
where $\left|eB\right|/(2\pi)$ is the density of gapless modes at the Fermi points, $|p_z|=p_F$. We note that our anomaly result is consistent with our CME result, Eq.~(\ref{cme}). Both vanish when matter disappears, $v_F\to0$ or $\mu\to m$. 
We also note that our calculations for the axial anomaly (and also for CME) are valid for arbitrarily small $B$, as it should be for anomaly, since only the gapless modes at the Fermi points of the LLL electrons contribute to anomaly. 

\section{Experimental setup}

\begin{figure}[t]
\centering 
\includegraphics[scale=0.35]{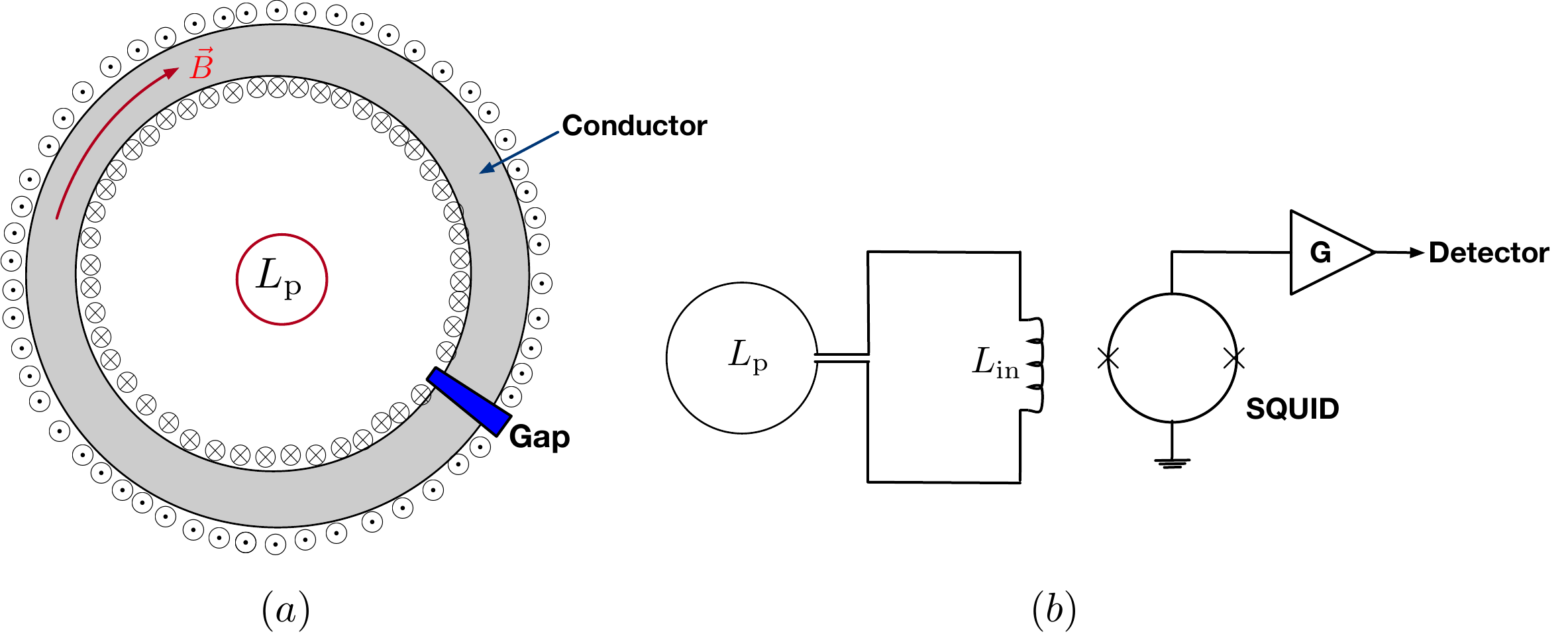}
\label{squid}
\caption{A simplified schematic of LACME.  A superconducting quantum interference device (SQUID) is used to pick up the magnetic flux, generated by the electric currents in the conductor inserted inside the gapped toroidal solenoid, shown in Fig.~2 (a). (Note that the toroidal solenoid has to have a gap in the superconducting solenoid as in ABRACADABRA, but not in the conductor inside the solenoid, for the Meissner current in the solenoid to close~\cite{Kahn:2016aff}). Fig.~2 (b) shows the schematics of the readout circuits and $L_{\rm p}$ denotes the pickup loop, placed at the center of the toroidal solenoid.}
\end{figure}

The experiment, schematically shown in Fig.~2, to measure the electric current due to ADM in medium will be similar to ABRACADABRA~\cite{Kahn:2016aff} and applicable to the cavity experiment~\cite{Sikivie:1983ip} as well. The main difference is however that one uses a conductor which will transport the electric charges along the external magnetic field without any supply of external voltages. With the superconducting pick-up loop, that measures the magnetic flux,  we then measure the electric current transported by the charge carriers of the conductor, generated by CME. 
Since the vacuum current due to the axial anomaly is always there, however, in our experimental setup, shown in Fig.~2\,(a),  the electric current inside the solenoid will be the sum of two currents, the vacuum current and the CME current,
\begin{equation}
\vec j=\left[C_{a\gamma\gamma}+4v_FC_e\right]\frac{\alpha}{2\pi f}\vec B\sqrt{2\rho_{\rm DM}}\cos\left(m_at\right)\,,
\label{current_a}
\end{equation}
where the anomaly coefficient $C_{a\gamma\gamma}$ defines the axion-photon coupling, $g_{a\gamma\gamma}=C_{a\gamma\gamma}\alpha/(2\pi f)$ with $\alpha$ being the fine structure constant and $\vec B=\mu_0(\vec H+\vec M)$ is the magnetic field inside the solenoid where a magnetized conductor with magnetization $\vec M$ is inserted~\cite{coey}.  The first term in Eq.~(\ref{current_a}) is the vacuum current coming from the axial anomaly, while the second term is the CME current, due to the medium. 
Similarly, one can apply this scheme to the cavity experiment~\cite{Sikivie:1983ip} for detecting axion dark matter. We propose to insert a conductor inside the cavity under an external magnetic field to generate the CME current, proportional to the axion-electron coupling $C_e$, on top of the vacuum current which always exists due to the anomalous axion-photon coupling or $C_{a\gamma\gamma}$ as shown in Eq. (\ref{current_a}).

\begin{figure}[th!]
\centering 
\includegraphics[scale=0.355]{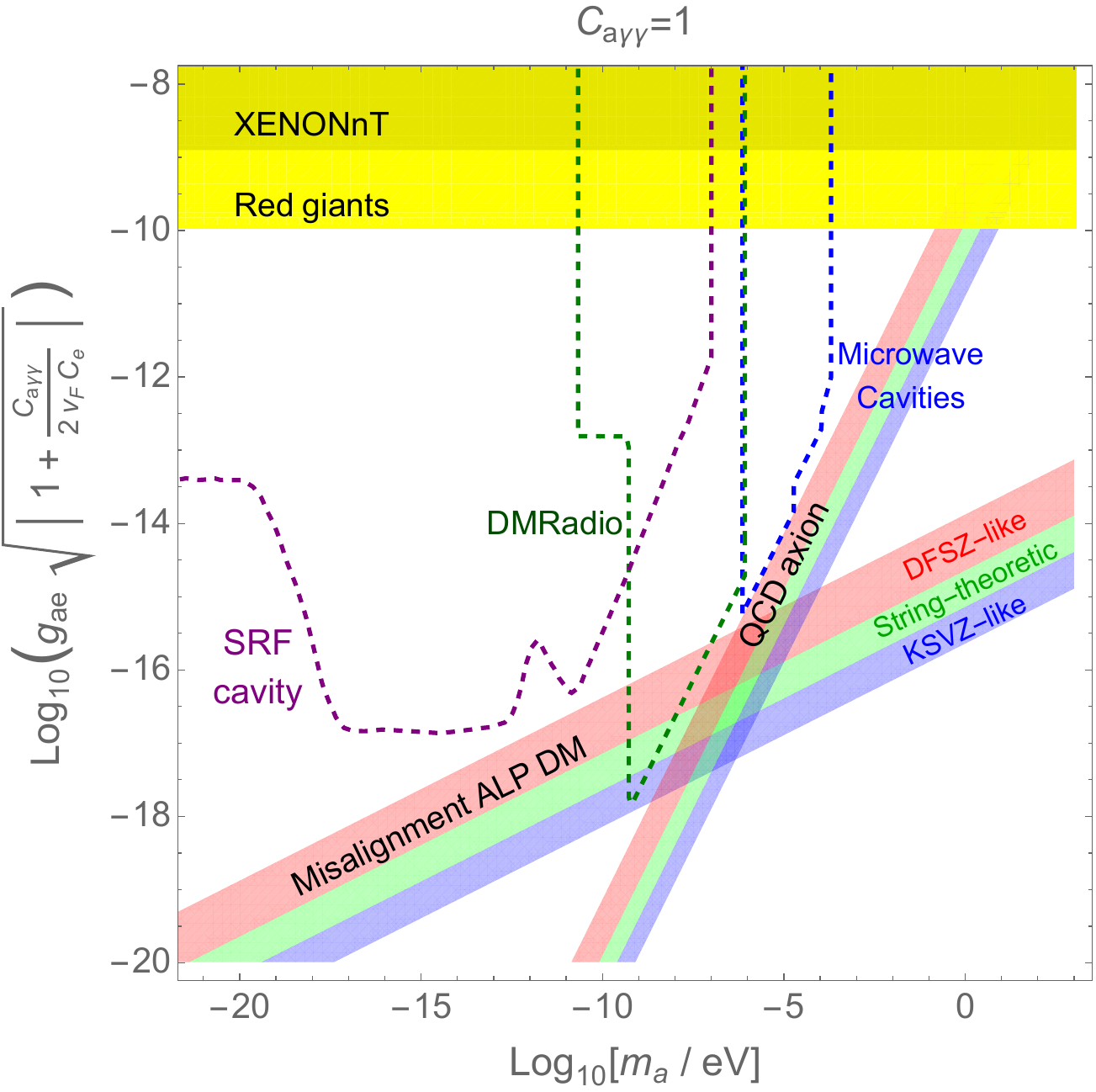}
\hspace{0.2cm}
\includegraphics[scale=0.34]{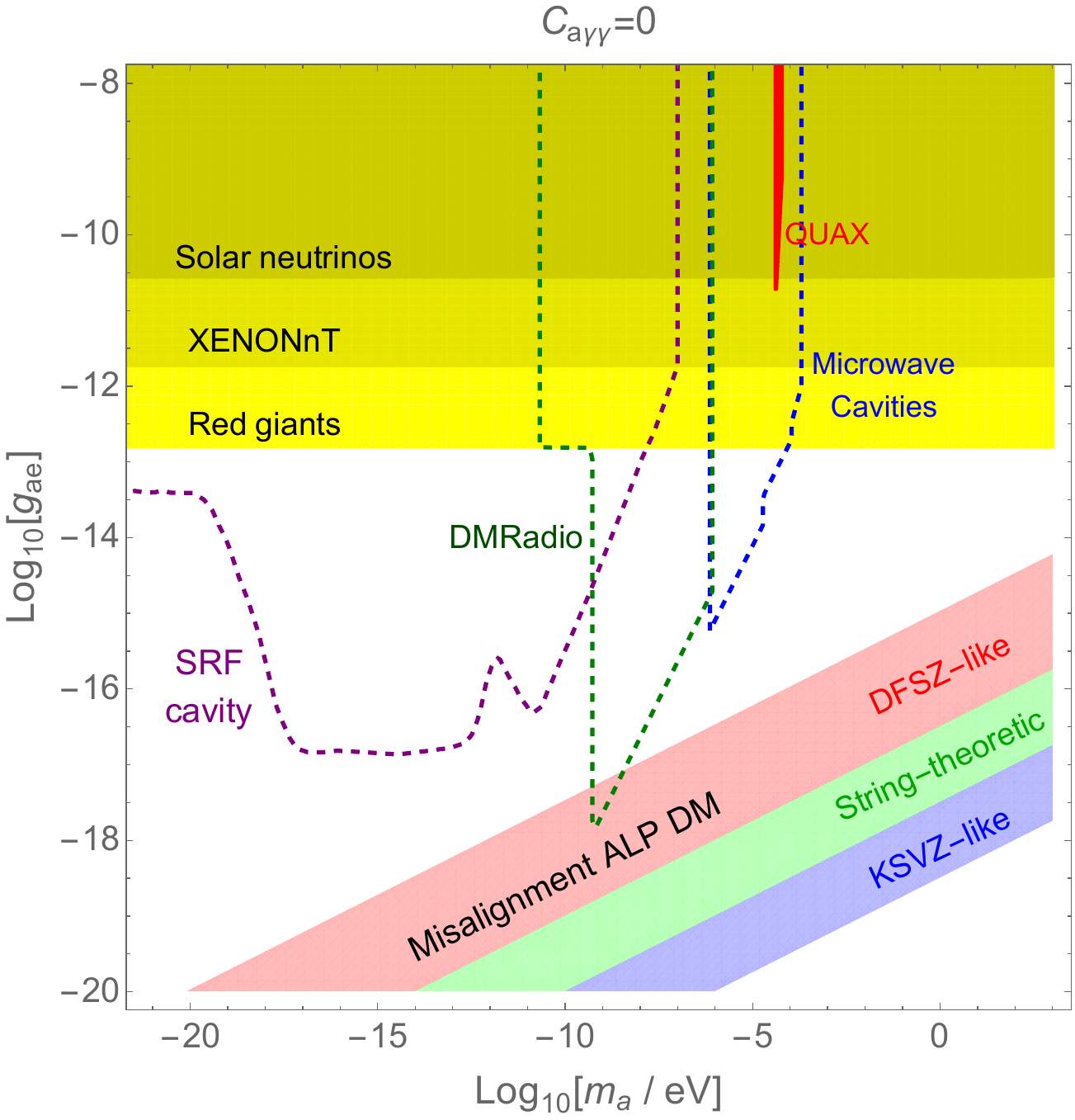}
\caption{The axion parameter space that can be probed by the axionic CME by repurposing the existing and proposed axion haloscopes such as microwave cavities \cite{ADMX:2020ote, Semertzidis:2019gkj, Beurthey:2020yuq, HAYSTAC:2023cam, Grenet:2021vbb, McAllister:2017lkb, QUAX:2023gop,  axionlimits}, DMRadio \cite{DMRadio:2022jfv}, and SRF cavity \cite{Berlin:2020vrk}, assuming $v_F=0.01c$, provided that LACME be adapted in those experiments with a conductor inserted in the haloscopes to generate the axionic CME current. Here $g_{ae} \equiv 2 C_e m_e/f$. The left plot is for the case that the axion DM couples to photons as in the case of QCD axion with a benchmark value $C_{a\gamma\gamma}=1$, while the right plot is for the case that the axion DM does not couple to photons unlike QCD axion. If $C_{a \gamma \gamma}\neq 0$, the axionic CME signals in the haloscopes depend on the product of $C_{a \gamma \gamma}$ and $C_e$ as in the left plot (see the text for details). The yellow shaded regions are excluded by the indicated astrophysical bounds \cite{Capozzi:2020cbu,Straniero:2020iyi, Gondolo:2008dd, XENON:2022ltv} (For the astrophysical bounds on the left, we give conservative bounds with $C_{a\gamma \gamma}/C_e \simeq 10^{4}$, a typical value for KSVZ-like axions), the red thick line is excluded by the QUAX experiment \cite{QUAX:2020adt}. The red-green-blue shaded bands correspond to the parameter space for the QCD axion and axion-like particle dark matter from the misalignment mechanism. 
    For the ALP dark matter, we assume that the ALP mass is independent of both temperature and the initial misaligned angle $\theta_i=1$. The red bands correspond to DFSZ-like models, the green bands for string-theoretic axions, and the blue bands for KSVZ-like models with the coefficient $C_e$ given in Eq. (\ref{Ce}).} \label{sensitivity}
\end{figure}

In order to experimentally separate the contribution of CME currents from the vacuum current, one has to repeat the experiment with and without a conductor inside the solenoid or the cavity. When applying our scheme to ABRACADABRA~\cite{Kahn:2016aff}, DMRadio~\cite{DMRadio:2022jfv}, microwave cavities   
such as \cite{ADMX:2020ote, Semertzidis:2019gkj, Beurthey:2020yuq, HAYSTAC:2023cam, Grenet:2021vbb, McAllister:2017lkb, QUAX:2023gop}, or the SRF cavity proposed in \cite{Berlin:2020vrk}, a major difference from those original experimental setups without any conductor inside would be that the noise budget has to include thermal noise caused by the conductor inserted inside the solenoid or the cavity. By the Dicke radiometer equation~\cite{Dicke:1946glx}, this thermal noise power is given by
\dis{
P_N^{\rm thermal} = k_B T \sqrt{\frac{\Delta \nu}{t_{\rm int}}}\,,
}
where $T$ is the effective noise temperature, $t_{\rm int}$ is the integration time for measurement, and $\Delta \nu$ is the bandwidth of signal frequency. Notice that the thermal noise power is determined by the quantities irrelevant of the details of the conductor.
We assume this thermal noise is the dominant additional source of the noise, coming from the inserted conductor.
One then finds that this additional noise is either comparable or subdominant to the dominant noises in the original experiments without the conductor inside such as DMRadio, the microwave cavities, and the SRF cavity.\footnote{In other experiments, e.g. the Search for Halo Axions with Ferromagnetic Toroids (SHAFT)~\cite{Gramolin:2020ict} and ABRACADBRA-10 cm~\cite{Salemi:2021gck} broadband experiments, the thermal noise from the inserted conductor is supposed to be bigger than the SQUID noise which is the dominant noise in such experiments without the conductor. For these experiments, one needs to reanalyze the signal-to-noise ratio including the thermal noise for the CME signal. We leave this for a future work.} Therefore, we can directly reinterpret the signal-to-noise ratio (SNR) previously estimated for the original experiments to our experimental setup. In DMRadio, the microwave cavities, and the SRF cavity, the signal power is proportional to the square of the current in Eq. (\ref{current_a}), and so with the inserted conductor
\dis{
P_S^{\rm conductor} \propto |C_{a\gamma\gamma}+4v_F C_e|^2/f^2. \label{PSwith}
}
In order to distinguish $C_e$-induced current from the $C_{a \gamma\gamma}$-induced one, one may repeat the experiment without the conductor and subtract the corresponding signal power $P_S^{\rm vacuum} \propto |C_{a\gamma\gamma}|^2$ from Eq. (\ref{PSwith}) :
\dis{
\left|P_S^{\rm conductor}-P_S^{\rm vacuum}\right| \propto \left(\frac{4 v_F C_e}{f}\right)^2 \left|1+\frac{C_{a\gamma\gamma}}{2v_F C_e}\right|\,.
}
If this signal difference is observed, one can determine the size of the axion-electron coupling, $C_e$ from the chiral magnetic effects. 
In Fig.~\ref{sensitivity}, we show the projected sensitivity for $g_{ae} \sqrt{\left|1+C_{a\gamma\gamma}/2v_F C_e\right|} \propto \left|P_S^{\rm conductor}-P_S^{\rm vacuum}\right|^{1/2}$ from each experiment by requiring the signal-to-noise ratio (SNR) $|P_S^{\rm conductor}-P_S^{\rm vacuum}|/P_N>1$, assuming the Fermi velocity $v_F=0.01c$ and the same noise budget as in the original experiment. 

Since the measurable quantity depends on the ratio $C_{a \gamma \gamma}/C_e$, we plot two cases in Fig.~\ref{sensitivity}, one in the left panel for a non-vanishing axion-photon coupling with a benchmark value $C_{a\gamma\gamma}=1$ and the other in the right panel for a vanishing axion-photon coupling $C_{a\gamma\gamma}=0$. For QCD axions, the axion-photon coupling is supposed to be non-vanishing with $C_{a\gamma\gamma} \sim {\cal O}(1)$ unless the models are fine-tuned \cite{DiLuzio:2016sbl}, while it is not necessarily non-vanishing for generic ALP dark matter. For this reason, we do not include the QCD axion bands for the case with $C_{a\gamma\gamma}=0$ in the right panel of Fig.~\ref{sensitivity}. On the other hand, we take the typical values of $C_e$ in Eq. (\ref{Ce}) for the QCD axion bands and the ALP dark matter bands in Fig.~\ref{sensitivity}. More precisely, we take the following benchmark values in the plot for the axion-electron coupling: 
\begin{equation}
C_e \simeq 
\begin{cases}
10^{-2}~\textrm{to}~\frac{1}{3} & \textrm{DFSZ-like models} \\
10^{-3} ~\textrm{to}~ 10^{-2} & \textrm{string-theoretic axions} \\
10^{-4} ~\textrm{to}~ 10^{-3} & \textrm{KSVZ-like models}.
\end{cases} 
\end{equation}

Fig.~\ref{sensitivity} clearly shows that the repurposed axion haloscopes with an inserted conductor (the ``LACME" scheme) can have promising sensitivity to reach the theoretically motivated parameter spaces of QCD axions and also ALP dark matter from the misalignment production mechanism. Although with the currently proposed axion haloscopes
only DMRadio may be able to probe some of ALP dark matter parameter space for the mass range from neV to 0.1 $\mu$eV, the LACME scheme in principle can exclude or detect the axion-electron coupling many orders of magnitude smaller than the astrophysical bounds and so strongly motivates improving the sensitivity of axion haloscopes beyond what is needed for probing the axion-photon coupling.

\section{Conclusion}

To conclude  we show that the chiral magnetic effect is realized in a Fermi liquid of electrons by demonstrating that electric currents generate spontaneously along the  external magnetic field if the medium is helicity imbalanced due to axion or axion-like dark matter. The CME is found to be suppressed by the Fermi velocity, in sharp contrast with the original result~\cite{Fukushima:2008xe}, which claims to be independent of the fermion mass, hence the Fermi velocity. We then propose to measure this spontaneous electric current in conductors to detect the axions or the axion-like particles. In our LACME experiment both CME current and the vacuum current exist. The measured current will be therefore proportional to $C_{a\gamma\gamma}+4v_FC_e$. In general the CME current is suppressed by a factor of Fermi velocity, compared to the vacuum current.  By repeating the experiment with and without the conductor inside the solenoid or the cavity of repurposed axion haloscopes, however, one can obtain the signal power proportional to $v_F C_{a\gamma \gamma} C_e+2v_F^2 C_e^2$,  which allows us to probe the axion-electron coupling $C_e$.
The magnitude of the CME current varies quite a lot, depending on UV models. Therefore precision measurements of the CME current can give us important information about the microscopic origin of QCD axion or ALP dark matter. We find that the LACME scheme is quite promising as it can marginally reach the theoretically expected parameter spaces for QCD axion or ALP dark matter with the currently proposed axion haloscopes. It motivates improving the sensitivity of axion haloscopes beyond the reach for the axion-photon coupling.  

\acknowledgments
This work was initiated at the 2022 CERN-CKC workshop on physics beyond the standard model, Jeju Island, June 2022 and one of us (DKH) is grateful to J. Foster and H. Kim for discussions during the workshop. We thank M. Cha, K. Choi,  Y.~C. Chung, H. Kang, Yoonseok Lee and C.~S. Shin for useful comments.
This work was supported by the National Research Foundation of Korea (NRF) grant funded by the Korea government (MSIT) (2021R1A4A5031460) (DKH, KSJ, DY), Basic Science Research Program through the National Research Foundation of Korea (NRF) funded by the Ministry of Education (NRF-2017R1D1A1B06033701) (DKH), and IBS under the project code, IBS-R018-D1 (SHI).


\end{document}